\documentclass[10pt]{iopart}
\usepackage{graphicx}
\usepackage[numbers,square,sort&compress]{natbib}
\usepackage[T1]{fontenc}
\usepackage[utf8]{inputenc}
\usepackage{natbib}
\usepackage{iopams}
\usepackage{graphicx}
\usepackage{xcolor}
\usepackage{hyperref}
\usepackage[normalem]{ulem}
\usepackage{txfonts}
\definecolor{myblue}{RGB}{0, 40, 170}
\definecolor{pzcol}{RGB}{50, 10, 160}
\hypersetup{
    unicode=true,          % non-Latin characters in Acrobat’s 
    pdftitle={Quantum chaos in Feshbach resonances of the ErYb system},    
    pdfauthor={Maciej Kosicki, Mateusz Borkowski, Piotr S. Żuchowski},     
    colorlinks=true,       % false: boxed links; true: colored links
    linkcolor=myblue,          % color of internal links
    citecolor=myblue,        % color of links to bibliography
    filecolor=myblue,      % color of file links
    urlcolor=myblue,           % color of external links
}

\begin{document}
\title[Quantum chaos in Feshbach resonances of the ErYb system]{Quantum chaos in Feshbach resonances\\of the ErYb system}

\author{Maciej B. Kosicki, Mateusz Borkowski\footnote{Corresponding author.}, Piotr S. Żuchowski}
\address{Institute of Physics, Faculty of Physics, Astronomy and Informatics, \\Nicolaus Copernicus University, Grudziadzka 5, 87-100 Torun, Poland}

\ead{pzuch@fizyka.umk.pl}

\begin{abstract}
We investigate ultracold magnetic-field-assisted collisions in the so far unexplored ErYb system. The nonsphericity of the Er atom leads to weakly anisotropic interactions that provide the mechanism for Feshbach resonances to emerge. The resonances are moderately sparsely distributed with a density of $0.1\,{\rm G}^{-1}-0.3\,{\rm G}^{-1}$ and exhibit chaotic statistics characterized by a Brody parameter $\eta \approx 0.5-0.7$. The chaotic behaviour of Feshbach resonances is accompanied by strong mixing of magnetic and rotational quantum numbers in near-threshold bound states. We predict the existence of broad resonances at fields $<300\,{\rm G}$ that may be useful for the precise control of scattering properties and magnetoassociation of ErYb molecules. The high number of bosonic Er-Yb isotopic combinations gives many opportunities for mass scaling of interactions. Uniquely, two isotopic combinations have nearly identical reduced masses (differing by less than $10^{-5}$ relative) that we expect to have strikingly similar Feshbach resonance spectra, which would make it possible to experimentally measure their sensitivity to hypothetical variations of proton-to-electron mass ratio. 
\end{abstract}

\section{Introduction}

One of the most important recent achievements in the field of ultracold quantum gases was the production of samples of highly magnetic atoms: chromium (Cr)~\cite{Griesmaier2005}, erbium (Er)~\cite{aikawa2012bec} and dysprosium (Dy)~\cite{lu2010trapping,lu2011strongly,lu2012quantum, Tang2015}. At large distances these atoms interact mostly through an anisotropic dipole-dipole interaction which decays more slowly than the usual van der Waals atom-atom interaction. As a result, a sample of highly magnetic atoms can exhibit a rich variety of phenomena distinct from the behaviour of systems of alkali-metal or alkaline-earth metal atoms. Highly magnetic atoms promise to be a key building block in accomplishing the vision of making a versatile toolbox for quantum simulations of condensed-matter physics and many-body phenomena~\cite{baranov2012condensed}. Some of these promises have already been fulfilled experimentally -- for example, the extended Bose-Hubbard model was recently realized experimentally using ultracold Er atoms confined to a 3D optical lattice~\cite{baier2016extended}. Quantum degenerate Er was used to study the deformation of a Fermi surface by the dipole-dipole interaction~\cite{aikawa2014observation} and the formation of quantum droplets~\cite{schmitt2016,chomaz2016quantum}. Apart from the applications in quantum many-body physics~\cite{DyManyBodyExample1,DyManyBodyExample2} the Dy atom is notably known for its role in studies of physics beyond the Standard Model~\cite{VanTilburg2015,budker1994}.

%\todo{Erbium is laser-coolable, with BEC, other lanthanides also successful}
%\todo{Dysprosium can be used for magnetometry and eEDM}
%\todo{Atomic dipole-dipole interactions are more useful than molecular for no external fields are needed to polarize atoms}
%\todo{Fermions are important  for quantum simuations}

Magnetic Feshbach resonances are one of the most important tools for controlling the interactions of ultracold atoms~\cite{chin2010}. 
In collisions of open-shell lanthanides with a non-zero orbital quantum number the spectrum is very dense: the anisotropic interaction potentials couple the magnetic Zeeman states with end-over-end rotational quantum number, which results in multiple bound states being coupled to the entrance $s$-wave channel. 
Such a pattern was initially predicted by Kotochigova and Petrov~\cite{kotochigova2011,petrov2012}, and confirmed experimentally for Er~\cite{aikawa2014observation}, and later for Dy atoms~\cite{baumann2014observation,maier2015emergence,maier2015broad,lucioni2018dysprosium}. 
The resonance pattern is not only dense, but also features the characteristics of {\em quantum chaos}, evidenced by correlations in the resonance spacings.

The positions of magnetic Feshbach resonances appear for field values at which the molecular  bound states cross the zero-energy threshold.
Quantum chaos emerges when the mean spacing between these bound states is comparable to the coupling strength between them. 
In this regime the couplings cause multiple avoided crossings to emerge which results in a spaghetti-like pattern of repelling energy levels and a highly congested spectrum of overlapping resonances. 
The resonances can not be interpreted individually, but their statistical behaviour can be analyzed instead (see pioneering work of Porter and coworkers eg. ~\cite{porter1956fluctuations,rosenzweig1960repulsion}) and the tools for such analysis include the Random Matrix Theory (RMT) developed by Wigner~\cite{wigner1993class} and Dyson~\cite{dyson1962}.  
% In the Er spectrum reported by Frisch et al. \cite{frisch2014qci},
For ultracold gases the first evidence of chaos in a Feshbach resonance spectrum was reported in Er dimer by Frisch et al. \cite{frisch2014qci}.
Since then, the chaotic character of bound states and Feshbach resonances in ultracold collisions was also found for several other systems, for example, in mixtures of Yb atoms in $^1$S$_0$ and $^3$P$_2$ states at large magnetic fields~\cite{green2016quantum}, and molecular collisions of Li atom with CaH and CaF~\cite{frye2016approach}.
On the other hand, the resonance spacings in collisions of Er and Li atoms in a magnetic field~\cite{gonzalez2015magnetically}, and if highly magnetic europium ($^7$S) with alkali metal atoms~\cite{zaremba2018magnetically} have both been shown to follow the non-chaotic poissonian distribution.

Here we investigate the interactions and Feshbach spectra between weakly anisotropic lanthanides and heavy spin-singlet atoms using ErYb as a prime example. Ytterbium, alongside Sr, is the most widely used spin-singlet atom with applications for optical clocks, quantum gases and quantum simulation~ \cite{Swallows2011,Bober2015,Takahashi2016,Gao2018}. 
ErYb should also be very attractive due to the fantastic mass-scalability of both Yb~\cite{Kitagawa2008, Borkowski2009} and Er (as shown here). Three of the bosonic Yb isotopes, $^{168}$Yb~\cite{Sugawa2011} and $^{170}$Yb~\cite{Fukuhara2007} and $^{174}$Yb~\cite{Takasu2003} form stable Bose-Einstein condensates, and several Fermi and Bose-Fermi and Bose-Bose gases~\cite{Fukuhara2007a, Fukuhara2009} have also been demonstrated experimentally.
Er and Yb atoms could form many isotopic mixtures, including two fermion-fermion mixtures with nearly equal masses and very close polarizabilities (hence, trapping properties). While an Er-Yb quantum degenerate mixture has not been achieved yet, it is clearly within the reach of present state-of-the-art techniques. Our predictions could be tested by forming a dual Mott insulator of Er and Yb atoms. Keeping unity filling for Er would prevent losses due to Er-Er Feshbach resonances.

We unveil the impact of anisotropy on the chaotic behavior in this system. The interaction between these two atoms is by far simpler than in case of the lanthanide dimer or lanthanide-alkali metal atom case and can be tackled by realistic {\em ab initio} calculations, which might help shed some light on much more complex systems involving lanthanides (mixed-lanthanide quantum gases, lanthanide+alkali metal atom). Given the progress in the field and possibility of mutual trapping of alike atoms the  results presented in this paper should hold for most combinations, including (Ho, Dy, Tm)+(Sr, Yb, Hg).

The paper is planned as follows. In the next chapter we describe the interaction between the erbium and ytterbium atoms focusing on details of {\em ab initio} calculations and their implications. Next we describe the methodology of solving the close-coupling equations in the magnetic field. In Section~\ref{sec:spectra} we analyse the mass-scaling behavior of the zero-field scattering lengths and show example ErYb magnetic Feshbach spectra. Next, we focus on the statistical properties of resonance positions showing evidence of quantum chaos. In Section~\ref{sec:anisotropy} we show that the weak anisotropy present in the interaction between Er and Yb atoms is responsible for its chaotic behaviour. Finally we look into the sensitivities of magnetic Feshbach resonances to small changes of the proton-to-electron mass ratio that, in ErYb, could be measured experimentally thanks to the existence of two isotopic combinations with nearly equal reduced masses. Section~\ref{sec:conclusion} concludes the paper.

\section{Interaction of Er and Yb atoms}

\begin{figure}%[!tbp]
    %\fbox{
    %\begin{minipage}{120mm}%{153mm}
    %    \flushleft
        \includegraphics[width=\textwidth]{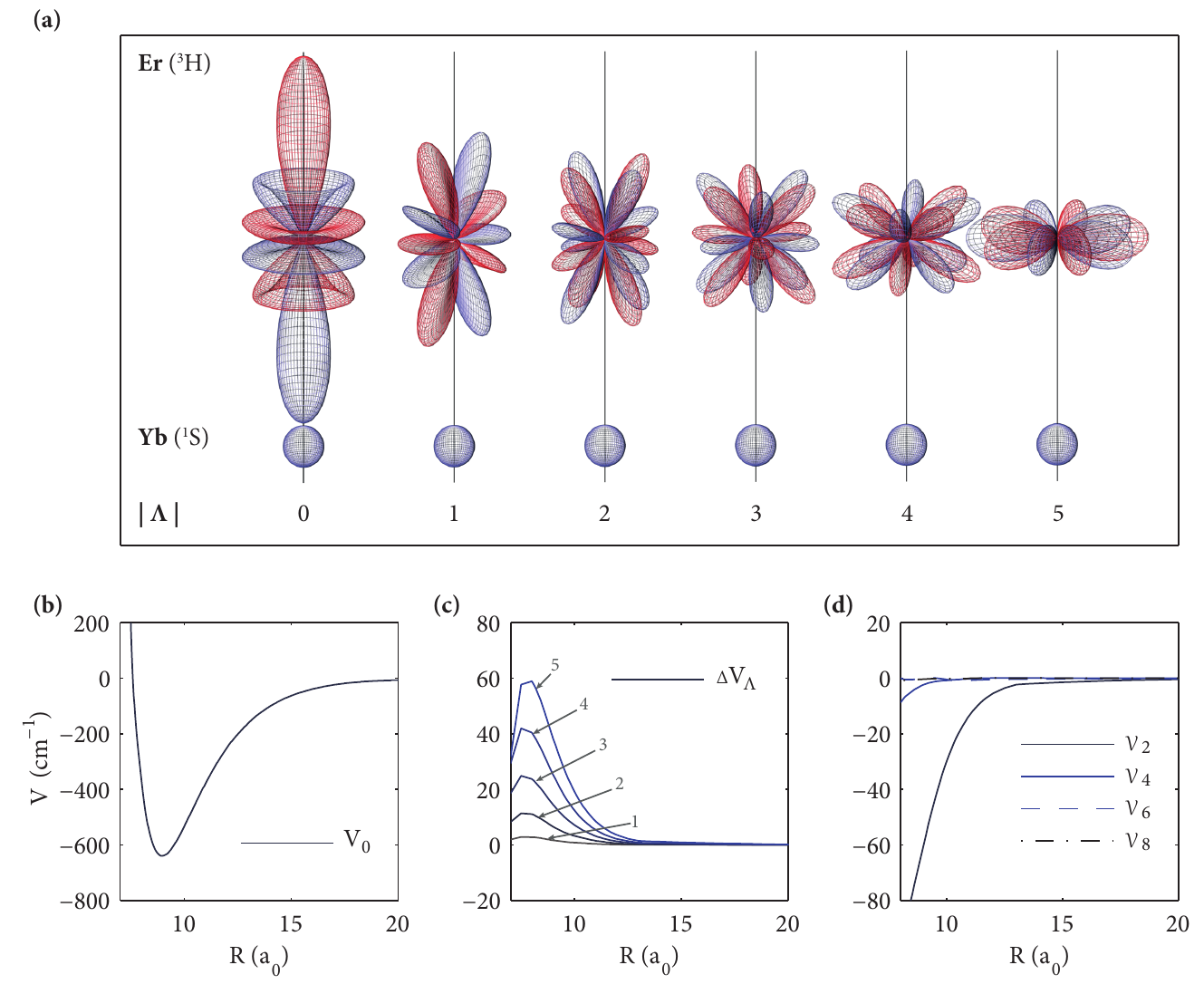}
        \caption{Anisotropic interactions in the ground-state Er($^3$H)+Yb($^1$S) system. 
                (a)~Schematic representation of the six $|\Lambda|$ states of the ground-state ErYb molecule.
                (b)~Broken-symmetry CCSD(T) X$^3\Sigma^-$ potential curve, $V_0$.
                (c)~Contributions ${\Delta}V_{\Lambda}$ from CASSCF potential energy curves to anisotropic states. These contributions are added to $V_0$ in order to compute $V_{\Lambda > 0}$.
                (d)~Anisotropic Legendre coefficients ${\cal V}_k$ for the short-range interaction potential based on the broken symmetry X$^3\Sigma^-$ state computed using the CCSD(T) method with added anisotropy. }
        \label{fig:PEC}
    %    \end{minipage}%}
\end{figure}

The erbium atom in its $^3$H$_6$ ground state is slightly anisotropic due to two missing electrons in the $4f$ shell screened by the doubly occupied $6s^2$ shell ($4f^{12}6s^{2}$). Since Yb has a spherical $^1$S ground state, the ground state ErYb molecule, shown schematically Fig.~1(a), inherits its quantum numbers from the erbium atom. The projection $\Lambda$ of the erbium atom's large orbital angular momentum ($l=5$) onto the interatomic axis gives rise to eleven states labeled by  $\Lambda=-l \ldots l$. The ($\Lambda=0$) $^3\Sigma^-$ state is unique while the remaining five $|\Lambda|>0$ states are doubly degenerate. 

Electronic structure calculations for such a heavy system as the ErYb molecule are challenging. Our general strategy follows Ref.~\cite{gonzalez2015magnetically} and involves the use of two complementary \emph{ab initio} methods. We first obtain the differences between complete active space self consistent field (CASSCF) total energies of all 11 states~\cite{CASSCF_ref1, CASSCF_ref2}. Since the CASSCF method itself does not reproduce the dynamic correlation properly, which results in a lack of the attractive dispersive interaction, we then add the calculated CASSCF anisotropies to the broken symmetry spin-restricted CCSD(T) lowest potential~\cite{Hampel:92, Knowles:94}. At long ranges we use van der Waals coefficients for isotropic part of the potential, $C_{6,0}$, and the leading anisotropy,~$C_{6,2}$.

For all of our electronic structure calculations we used the \textsc{Molpro} quantum chemistry code~\cite{MOLPRO_brief:2015}. Quasi-relativistic effective core potentials ECP28MWB and ECP60MDF for the first 28 and 60 electrons in Er and Yb atoms, respectively~\cite{ECP_Er,ECP_Yb}, were used to replace the core electrons. The ECP28MWB gaussian basis set for Er included the $[14s13p10d8f1g]$  functions with uncontracted \textit{s} and \textit{p} functions, and an additional $g=4.0$ exponent. For Yb, we used the $[7s7p6d8f1g]$ basis set in which the \textit{f} and \textit{g} functions were added from the Er atom. The basis set was also augmented by midbond functions to improve the accuracy in the potential well region. Finally, the description of both atoms was modified by adding one shell of even-tempered diffuse functions per orbital angular momentum.  
 
In their early studies of interactions of the lanthanides, Buchachenko and coworkers~\cite{buchachenko2007interactions,buchachenko2006van,buchachenko2006ab} demonstrated that the interaction has a peculiar character: the atomic anisotropy is suppressed due to the shielding by the doubly occupied outermost electronic shell. Systems like Tm+He and Er$({}^3$H$_6)$+Li$({}^2$S${}_{1/2})$ have been previously studied using
% the complete active space self-consistent field method %już zdefiniowane wcześniej
CASSCF~\cite{buchachenko2006van, gonzalez2015magnetically, CASSCF_ref1, CASSCF_ref2}.  This method preserves the symmetry associated with the projection $\Lambda$ of the orbital angular momentum $l$ on the interatomic axis; therefore, it allows to compute potentials corresponding to pure $|\Lambda|$ states. In our study of the Er($^3$H$_6$)+Yb($^1$S$_0$) system, we have performed CASSCF calculations with an active space including $4f6s6p$ orbitals for the Er atom, and $6s6p$ orbitals for the Yb atom. We carefully monitored the orbital symmetry of the optimized states by controlling the projection of orbital angular momentum on the molecular axis. In Fig.~\ref{fig:PEC}(c) we show the differences between $|\Lambda|>0$ potentials with respect to the state with the lowest energy, $^3\Sigma^-$ state. The potential anisotropies corresponding to $|\Lambda|>0$ states range from a few to several dozens of cm${}^{-1}$ and increase with $|\Lambda|$.  

The CASSCF interaction potentials do not reproduce the dynamic correlation correctly which results in an absence of dispersion forces and, as a consequence, incorrect long range asymptotics. To fix this we performed spin-restricted CCSD(T) calculations, which for similar systems provides few-percent accuracy~\cite{Smith2014}. However, we were only able to converge the CCSD(T) calculations for four states (instead of all eleven), and we were not able to control the $|\Lambda|$ quantum number in them. Hence we used the interaction energy for the lowest state obtained with CCSD(T), shown in Fig.~\ref{fig:PEC}(b), combined with the energy differences obtained with CASSCF to model the atomic interaction. 

As a sanity check, we have also performed broken-symmetry MP2 calculations based on the restricted open-shell Hartree-Fock determinant, as well as CASPT2 calculations~\cite{CASPT2}. The spectroscopic parameters of potential energy curves obtained with CCSD(T), MP2 and CASSCF methods are shown in Table~\ref{tab:PEC}. The spectroscopic constants predicted by all three methods are quite similar. The dissociation energy for ErYb obtained with broken-symmetry CCSD(T) is 639~cm${}^{-1}$. This value can be compared with experimental estimates of the dissociation energy for Yb$_2$ of 738~cm$^{-1}$~\cite{Borkowski2017} and RbYb of 785~cm$^{-1}$~\cite{Borkowski2013}. In the latter case, \emph{ab initio} calculations based on the ECP60MDF core potential as reported in Ref.~\cite{Borkowski2013} predicted a well depth of 704 cm$^{-1}$, hence we expect our result could be underestimated by about 10\%.

At long ranges the interaction between Er and Yb atoms is dominated by the van der Waals interaction potential given by $V(R,\theta)=-C_{6,0}R^{-6} -C_{6,2}R^{-6} P_2(\cos \theta)$. Higher anisotropy terms decay much faster with distance (at least $R^{-8}$) and are thus ignored in this work. We have estimated the $C_{6,0}$ coefficient by means of a combination rule proposed by Tang~\cite{Tang1969} where  $C_{6,0}$ can be obtained from static polarizabilities of atoms $\alpha_{\rm Er}$, $\alpha_{\rm Yb}$ and homonuclear coefficients $C_{6,0, {\rm Er}_2}$ and $C_{6,0, {\rm Yb}_2}$:
\begin{equation}
C_{6,0} =  \frac{2 \alpha_{\rm Er} \alpha_{\rm Yb} C_{6,0, {\rm Er}_2} C_{6,0,{\rm Yb}_2}}
{( \alpha_{\rm Er}^2 C_{6,0,{\rm Yb}_2} + \alpha_{\rm Yb}^2 C_{6,0,{\rm Er}_2})}\,.
\end{equation}
The anisotropic $C_{6,2}$ coefficient can be evaluated by scaling the isotropic $C_{6,0}$ by the ratio of the erbium atom's tensor and scalar polarizabilities:
\begin{equation}
C_{6,2} = \frac{\alpha_{2, {\rm Er}}}{\alpha_{0, {\rm Er}}} C_{6,0} \,.
\end{equation}
The necessary atomic polarizabilities and homonuclear $C_{6,0}$ factors can be found in Refs.~\cite{Becher2018, Safronova2012, Lepers2014, Kitagawa2008, Chu2007}. For ErYb we find $C_{6,0}=1835\, E_h a_0^6$ and $C_{6,2}=33.62\,E_h a_0^6$ (the Hartree energy $E_h\approx4.359744650(54)\times 10^{-18}$~J and Bohr radius $a_0\approx5.2917721067(12)\times10^{-11}$~m are the atomic units of energy and distance).

\begin{table}
	\caption{   
	 Spectroscopic constants for the lowest interaction potential of the Er$({}^3$H$)$+Yb$({}^1$S$)$ system including equilibrium distance $R_e$, well depth $D_e$, rotational constant $B_{\rm rot}$ and harmonic frequency $\omega_0$. The latter two correspond to the $^{166}$Er+$^{174}$Yb isotopic pair composed of the two most abundant isotopes. 
	}
	\flushright
	 \begin{tabular}{ccccccc}
	    \br
	     method & $R_e$ (a${}_0$) 	& $D_e$ (cm${}^{-1}$) &$B_{\rm rot}$  (cm${}^{-1}$) & $\omega_0$ (cm${}^{-1}$)  \\
	     \mr
CCSD(T) &	8.96&  639 & 0.0025 &	12.43 \\
MP2     &	8.87&  763 & 0.0025 &	12.07 \\
CASPT2  &	8.87&  671 & 0.0025 &	12.98 \\
	  	    \br
	 \end{tabular}
	\label{tab:PEC}
\end{table}

\section{Quantum scattering calculations}
While Yb, in its $^1$S$_0$ ground state, is a structureless atom, Er has a highly magnetic $^3$H ($l=5$, $s=1$) ground state which splits into three fine sublevels $j=4,5,6$ with the $^3$H$_6$ state being the lowest in energy. Bosonic isotopes of both Er and Yb also lack hyperfine structure. Since the fine splitting of more than 6000~cm$^{-1}$ between the $^3$H$_6$ ground state and the closest excited state, $^3$H$_5$, is an order of magnitude larger than the potential depths, we carry out our scattering calculations for just the lowest $^3$H$_6$+$^1$S$_0$ asymptote. Our basis spans the Hilbert space of 
\begin{equation}
	\left| j(ls) m_j \right \rangle %\otimes
	\left| L M_L \right>,
\end{equation}
molecular states where $j=6$, $l=5$, $s=1$, $m_j=-j,...,+j$. %In a magnetic field $j$ is not conserved, but its projection onto the space-fixed $z$ axis, $m_j$, is. 
We assume $s$-wave ($L=M_L=0$) scattering in the lowest Zeeman sublevel $m_j$. The rotational (end-over-end) quantum number only takes even numbers $L=0,2,\ldots$ because the potential anisotropy only couples partial waves differing by an even number. The projection of $L$ on the laboratory axis $M_L = -L,-L+1,\ldots,L$ as usual, but is subject to the restriction that during a collision the projection of the total angular momentum $m_j+M_L$ is conserved. The collision Hamiltonian
\begin{equation}
\hat{H}= -\frac{  \hbar^2}{2 \mu} R^{-1}  \frac{ d^2}{d R^2}R  +   \frac{  \hbar^2  \hat{L}^2 }{2 \mu R^2} +  \hat{\cal V} + \hat{H_Z},
\end{equation}
consists of, respectively, kinetic and rotational energy, the atomic interaction potentials $\hat {\cal V}$ and Zeeman interaction of the isolated Er atom with the external magnetic field $\vec B$, $\hat{H}_{Z} = \mu_B  B g_j \hat{j}_z$. The Land\'e g-factor $g_j=1.166$; $R$ and $\mu$ denote the interatomic distance and reduced mass. 

It is standard procedure for the interaction potential operator  $\hat{\cal V}$ to be expanded in Legendre polynomials~\cite{Callaway1965,Reid1969,Reid1973}:
\begin{equation}
\hat{\cal V}= \sum_k  P_k(\cos \theta) {\mathcal{V}}_k (R),
\end{equation}
where
$k = 0, 2, ...2l$.
The matrix elements of the angular part $P_k(\cos \theta)$ in the $j$-coupled basis set are
%\begin{widetext}
\begin{eqnarray}
\label{Vcpled}
& \fl \sum_k  \langle L M_L |  j(ls) m_j   | P_k(\cos \theta)  | j'(ls) m'_j | L' M'_L \rangle {\mathcal{V}}_k = \nonumber \\
    \sum_k & (-1)^{ s +j + j' + k  + m_k - m_l - m_j} (2l+1) \sqrt{(2j+1) (2j'+1)(2L+1)(2L'+1)}
 \nonumber \\
	&\times \left\{ \begin{array}{ccc}  l & j  &  s \\ j' & l  & s \end{array} \right\}
	\left( \begin{array}{ccc}  j & k &  j' \\ -m_j & m_k & m'_j \end{array} \right)
	\left( \begin{array}{ccc}  L & k &  L' \\ -M_L & -m_k & M'_L \end{array} \right) \nonumber \\
	&\times \left( \begin{array}{ccc}  l & k &  l \\ 0 & 0 & 0 \end{array} \right)
	\left( \begin{array}{ccc}  L & k &  L' \\ 0 &0  & 0 \end{array} \right) {\mathcal{V}}_k,
\end{eqnarray}
%\end{widetext}
where $m_k = M_L - M'_L$, symbols in parentheses and curly braces are 3-$j$ and 6-$j$ symbols~\cite{Krems2003}.
The potential matrix changes the $m_l$ quantum number such that $m_l + M_L$ is conserved and the mixing between different $m_j$ exists for each element with nonzero $k$. To be more precise, $k>0$ directly couples the channels with $\Delta m_j=\pm 1, \pm 2, \ldots, \pm k$ at the cost of $M_L$ quantum number. The higher anisotropies $k=2,4\ldots$ were obtained using the following transformation~\cite{Krems2004, Maykel2013}
  \begin{equation}
{\mathcal{V}}_k(R) = \frac{2k+1}{2l+1} \sum_{\Lambda = -l}^l \frac{\langle l \Lambda k 0| l \Lambda \rangle }{ \langle l 0 k 0| l 0 \rangle } V_{\Lambda} (R).
\end{equation}
Fig.~\ref{fig:PEC}(d) shows the anisotropic Legendre coefficients for erbium-ytterbium interaction for $k=2\ldots8$.

All calculations with the collisional Hamiltonian were done with the close-coupling (coupled channel) code implemented by Hutson and LeSueur in \textsc{bound} (for the calculations of bound states in the magnetic field), \textsc{field} (for converging the value of magnetic field at which the bound state crosses the threshold)~\cite{BoundField}, and \textsc{molscat}~\cite{molscat_new} programs (for scattering length). We used the hybrid log-derivative-Airy method of Alexander and Manopoulos~\cite{Alexander1987} with a fixed step size of 0.002~$a_0$ for $R$ between $6.2\,a_0$ and $30\,a_0$ and a variable step size for $R$ between $30\,a_0$ and $800\,a_0$. The collision energy was $E/k_B = 10^{-8}\,$K.

 The anisotropic interaction potential $\hat{\cal V}$ couples the initial $s$-wave ($L=0$) scattering channel to  many  closed channels that correspond to higher partial waves.  A practical calculation, however, can only include a finite number of partial waves, and the resulting spectrum of Feshbach resonances will depend on the selected basis set. Generally, as the number of included partial waves is increased, the total number of Feshbach resonances also grows. The width, however, of resonances originating from higher partial waves quickly becomes vanishingly small, as the coupling of the $s$-wave to higher partial waves becomes weaker, and the density of states of  closed channels at the  threshold decreases as $L$ quantum number increases. We have found that the number of resonances with an appreciable width -- larger than $0.01$~G -- saturates quickly and for the remainder of the paper we  calculated the properties of resonances which include $L$ up to 20:
a comparison with calculations performed for maximum $L=50$ for the magnetic fields up to 100 G shows that the calculations with basis set $L=50$ produces 10 more resonances (48 compared to 38) but all new resonances have a width smaller than 0.01 G. 

% \todo{ Maciej jeszcze potwierdzi jaka jest maksymalna szerokosc rezonansu nie ujetego w LMAX=20 a obecnego w LMAX=50}  
% }
% MK: Zmiana rozmiaru bazy z $L_{max}=20$ do $L_{max}=50$ powoduje, ze w polu do 100 G pojawiaja sie dodatkowe rezonanse, z ktorych najszersze sa rzedu 10${}^{-2}$ G. 

%. One might ask on what value one should truncate $L$ so that the pattern of resonances is properly converged.
%As we increase the basis set new resonances appear, but their width decreases with increasingly higher $L$, which is due to weaker coupling of $s$-wave to high $L$ through the interaction potential.
%In the Tab.~\ref{tab:convergence} we demonstrate the properties such as the average spacing of resonances and their number for basis sets with increasing $L_{\rm max}$ parameter, for the magnetic field up to 100 G. While the number of resonances indeed increases with $L_{\rm max}$,
%the number of resonances larger than 0.1 G saturates very quickly.  Interestingly, the narrow resonances need more partial waves to converge their position, while for the broad ones, which are dominated by channels with the low $L$ the exact position of resonance converges very fast. 
%\todo{we show convergence using two example resonances, wide resonance converges quickly, narrow converges slowly}
%\todo{resonance positions also depend on ell max because of strong mixing between partial waves dominated by V2}%\todo{the number of resonances delta b>0.2 g converges at ell max 20}

\section{Zero-field scattering lengths and Feshbach spectra \label{sec:spectra}}
%\pzuch{Perhaps we should start from bound states at zero field and reduced mass scaling pattern }
\begin{figure}
    %\fbox{
    %\begin{minipage}{153mm}
    %    \flushright
        \includegraphics[width=\textwidth]{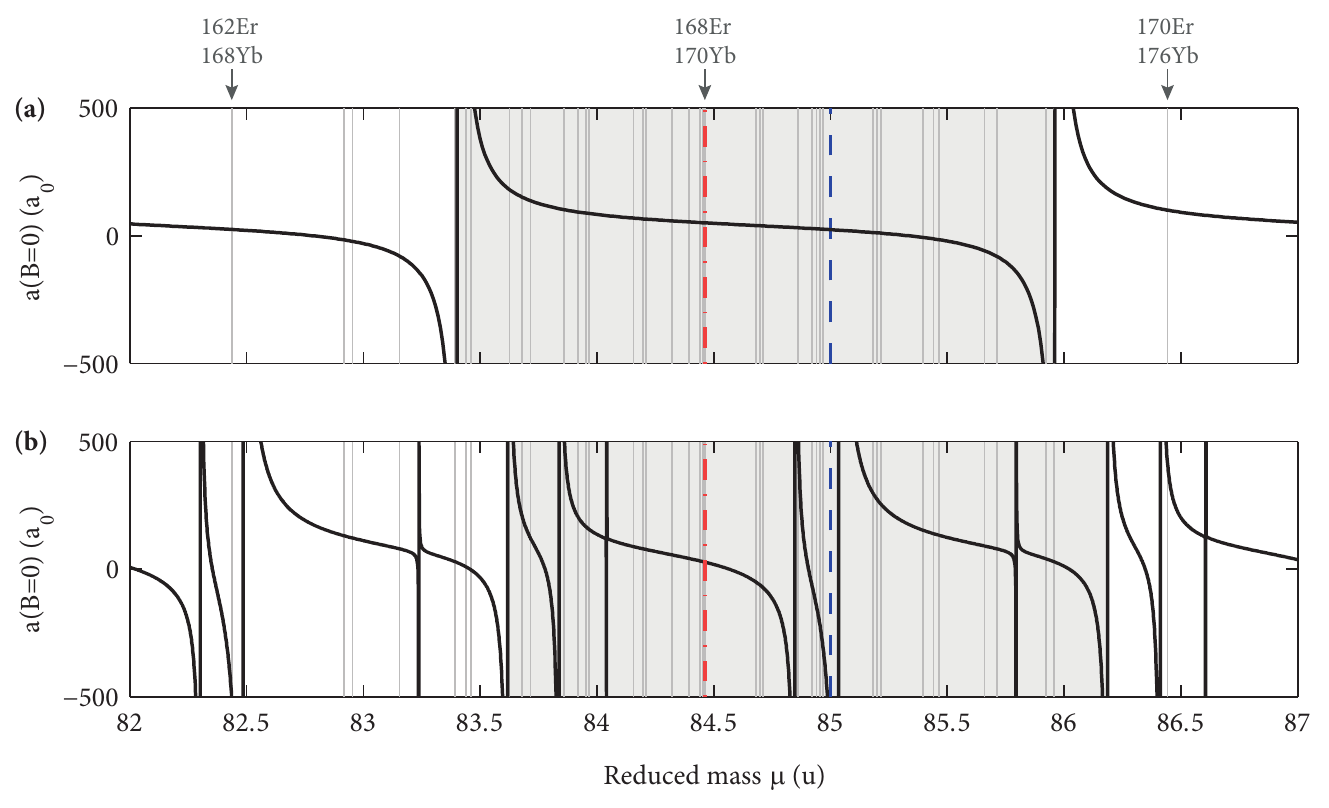}
        \caption{\textbf{Mass scaling of zero-field scattering lengths.} 
            (a) Zero-field scattering lengths for the isotropic potential only.
            (b) Scattering lengths for fully anisotropic interaction. 
            The shaded area corresponds to a full cycle of scattering lengths.
            The light grey vertical lines correspond to reduced masses of ErYb isotopic combinations. The red and blue lines correspond to the respective ``off-resonance'' and ``on-resonance'' cases discussed later in text.
        }
        \label{fig:zerofield}
    %    \end{minipage}}
\end{figure}

%Let us first focus on the  behaviour of scattering length in field-free case. 
Due to the large reduced mass of the Er+Yb system, our isotropic potential $V_0$ supports a large number of bound states, between 64 for the lightest and 66 for the heaviest isotopic combinations. In absence of anisotropy, the mass variation of the zero-field \mbox{$s$-wave} scattering length [Figure \ref{fig:zerofield}(a)] follows the well known periodic behaviour, $a(B=0) \approx \bar a \left[1-\tan \left(\phi-3\pi/8 \right)\right]$, where $\phi=\hbar^{-1}\int_{R_t}^{\infty} \sqrt{-2\mu V(r')} dr'$ is the WKB phase integral and $\bar a = 2^{-3/2}[\Gamma(3/4)/\Gamma(5/4)](2\mu C_6/\hbar^2)^{1/4}\approx 73\,a_0$ is the ``mean scattering length''~\cite{Gribakin1993}. 
The range of reduced masses covered by all isotopic combinations (vertical lines in Figure \ref{fig:zerofield}(a)) covers approximately one and a half cycle of $a$.

Figure \ref{fig:zerofield}(b) shows the scattering length once the anisotropic interaction is switched on. The anisotropic $V_k$ terms couple the $s$-wave to channels corresponding to higher partial waves causing additional (Feshbach) resonances to emerge. As result we obtain a complex, yet periodic structure. The resonance pattern, as we will show, is in itself chaotic. The periodicity, however, may stem from the near-threshold physics being governed mainly by the long range van der Waals interaction, and the WKB phase, $\phi$. The length of this cycle, of 2.6~u, aligns perfectly with the length of the cycle in $V_0$ alone.

Since we cannot predict the scattering lengths \emph{ab initio}, we instead study the Feshbach spectra for two representative cases, which we shall dub ``off-resonance'' and ``on-resonance''. The ``off-resonance'' case, shown in Fig. \ref{fig:zerofield} in red, corresponds to a zero-field $s$-wave scattering length close to $\bar a$, which should be representative for most isotopes. For the much fewer isotopes that lie on-resonance, we also calculate a spectrum for a reduced mass that for our model coincides with a resonance (Fig. \ref{fig:zerofield}, blue).

\begin{figure}
    %\fbox{
    %\begin{minipage}{153mm}
        %\flushright
        \includegraphics[width=\textwidth]{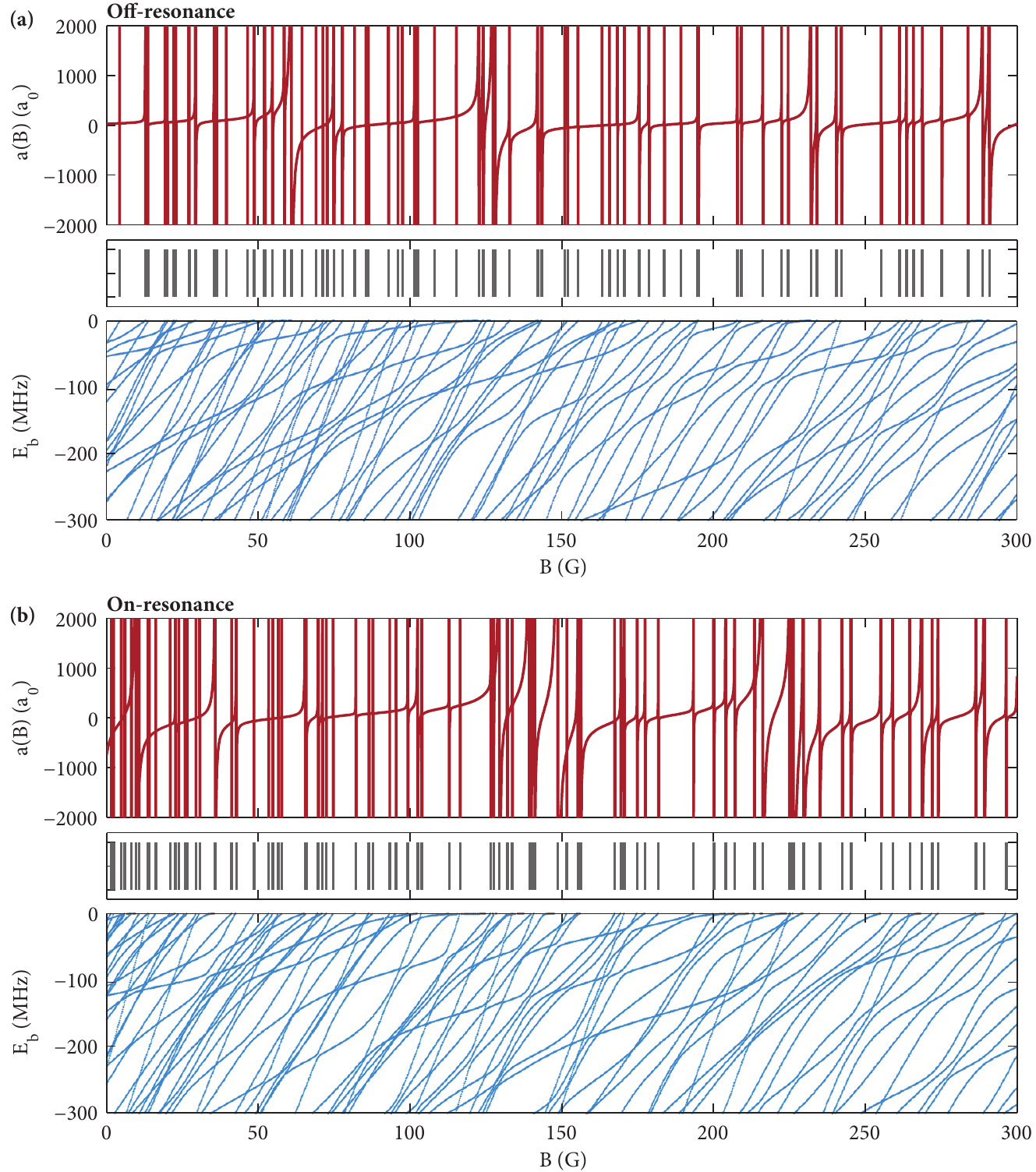}
        \caption{ Er+Yb magnetic Feshbach resonances: 
        variation of the scattering length (top panel),    positions (middle panel with grey stripes), and near-threshold bound state positions.
                 (a) Feshbach spectrum for a zero-field scattering length close to $\bar a$ (``off-resonance''). 
                 (b) Feshbach spectrum for a large negative scattering length (``on-resonance''). 
        }
        \label{fig:resonancespectra}
    %    \end{minipage}}
\end{figure}

\begin{figure}
    %\fbox{
    %\begin{minipage}{153mm}
        %\flushright
        \includegraphics[width=\textwidth]{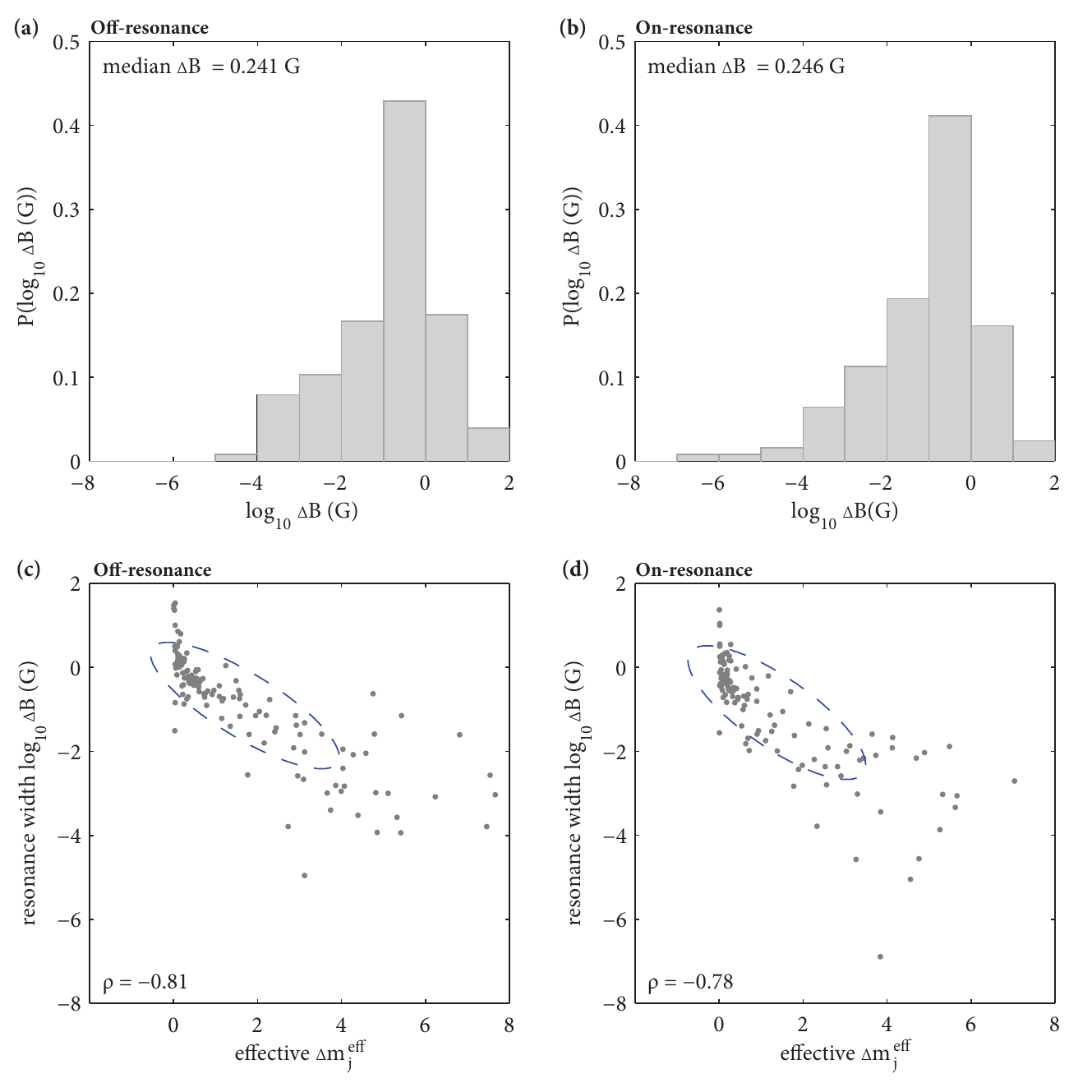}
        \caption{Widths of Er+Yb magnetic Feshbach resonances.
                (a-b) Histogram of log resonance widths for the resonances matching to the Breit-Wigner profile for the off-resonance and on-resonance datasets.
                (c-d) Correlation between log resonance widths and effective $\Delta m_j^{\rm eff}$ (see text for details). Covariance ellipses are shown to guide the eye.
        }
        \label{fig:widths}
    %    \end{minipage}}
\end{figure}

Figure \ref{fig:resonancespectra} shows the magnetic Feshbach spectra, resonance positions and near-threshold bound states as a function of the magnetic field $B$. For clarity we only show the range of $B \in [0,\,300]$~G, although we have calculated the spectra up to 1000~G. The resonance spectrum is dense, with many overlapping resonances. The resonance density is significantly higher at small fields: the full 1000~G scan yields 168 and 173 resonances for the ``off-resonance'' and ``on-resonance'' cases, respectively, but as much as 31 and 38 are contained within the first 100~G. While the overall densities of the spectra between the two samples are similar, no individual features from one spectrum can be discerned in another. The complexity of the Feshbach spectrum originates from the chaotic behavior of near-threshold bound states characterised by multiple avoided crossings caused by the anisotropic interaction.

Recently, Frye~\emph{et al.}~\cite{frye2017} published a procedure for extracting parameters, namely the positions $B_{\rm res}$, widths $\Delta B$, and background scattering lengths $a_{\rm bg}$, of Feshbach resonances that conform to the Breit-Wigner profile,
\begin{equation}
    a(B) \approx a_{\rm bg} \left( 1- \frac{\Delta B}{B-B_{\rm res}}\right)\,. \label{eq:breitwigner}
\end{equation}
We have found it applicable to the majority of the resonances in our spectra. Histograms of the resulting widths (in log scale) are shown in Figures~\ref{fig:widths}(a)~and~(b). In both cases the vast majority of resonances is narrower than 1~G and about 10\% is narrower than 1~mG. The respective median widths for the ``off-resonance'' and ``on-resonance'' datasets are nearly identical: $\Delta B = 0.241~{\rm G}$ and $\Delta B = 0.246~{\rm G}$. The distribution quartiles are qualitatively similar: the middle $50\,\%$ of resonances are contained within the ranges of  $\left<0.026, 0.800\right>~{\rm G}$ and $\left<0.126, 0.659\right>~{\rm G}$, respectively.

Narrower resonances originate from higher partial waves than broad resonances. We illustrate this by evaluating the differences in magnetic moments, $\Delta m^{\rm eff}_j g_j \mu_B$, between the initial scattering channel and the bound states the resonances originate from. Since the interaction potential only couples channels with matching $m_j+m_l$, $\Delta m_j$ determines $\Delta m_l$, which in turn provides indirect information on the partial wave composition of the bound state. Bound states are, in principle, mixtures of many partial waves. While we can not determine the actual composition of a bound state in terms of pure $m_j$ states, we can determine its magnetic moment $\Delta m^{\rm eff}_j g_j \mu_B = \partial E/\partial B$. In practice, we evaluate this derivative by probing the resonance position upon small changes to threshold energy $E$. Fig.~\ref{fig:widths}(b) shows log resonance widths as a function of $\Delta m^{\rm eff}_j$. The correlation coefficients $\rho=-0.81$ and $\rho=-0.78$ for the respective ``off-resonance'' and ``on-resonance'' datasets show clear correlation between (log$_{10}$) resonance widths and $\Delta m_j^{\rm eff}$.

%\todo{ na koncu jezyka mam: Interestingly we found almost similar pattern of resonances using only the $V_2$ potential, which indicates that the couplings between closed-channels are more important than direct couplings between scattering state and closed channels through higher order anisotropies $V_4$, $V_6$ etc.  }  

%\clearpage
\section{Resonance statistics and evidence of quantum chaos\label{sec:chaos}}

\begin{figure}%[!tbp]
    %\fbox{
    %\begin{minipage}{153mm}
    %    \flushright
    %    \vspace{6mm}
        \includegraphics[width=\textwidth]{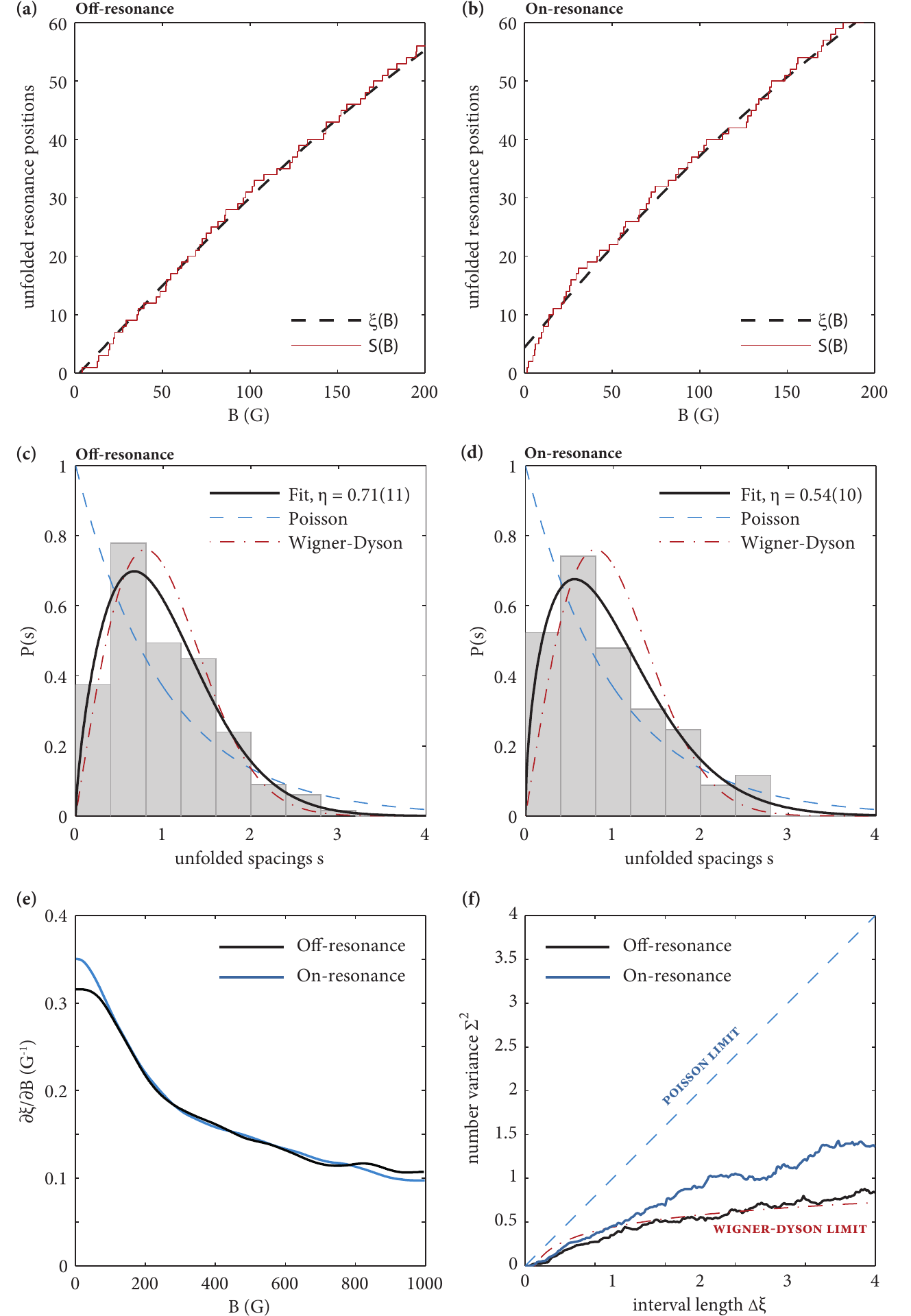}
        \caption{Statistical properties of ErYb magnetic Feshbach resonance spectra.
                 (a-b) Unfolding the resonance positions: $S(B)$ is the staircase function, $\xi(B)$ is the field-dependent ``resonance number'' for our example ``off-resonance'' [Fig.~\ref{fig:resonancespectra}(a)] and ``on-resonance'' [Fig.~\ref{fig:resonancespectra}(b)] spectra, respectively. Note that while the spectra were calculated for magnetic fields of up to $B=1000\,{\rm G}$, the figures are zoomed in for clarity.
                 (c-d)~Histograms of nearest-neighbor spacings (with respect to local resonance density). Best fit Brody distribution, as well as Poisson and Wigner-Dyson distributions are also shown.
                 (e)~Local resonance density as a function of magnetic field $B$. 
                 (f)~Number variances $\Sigma^2(\Delta\xi)$ as a function of interval length $\Delta\xi$. See Section~\ref{sec:chaos} for details. 
        }
        \label{fig:unfolding}
    %    \end{minipage}}
\end{figure}

Our calculated magnetic Feshbach spectra exhibit a chaotic behaviour. To show this, we quantify the resonance statistics using two standard measures of chaos: the Brody parameter $\eta$, which probes the propensity of adjacent resonances to repel each other~\cite{Brody1973}, and resonance number variance, a measure of long range resonance correlations~\cite{Weidenmuller2009}.

%For small magnetic fields $B$ our calculated magnetic Feshbach resonance spectra tend to be denser than at large $B$. Our ``off-resonance'' and ``on-resonance'' spectra (Figs.~\ref{fig:resonancespectra}(a) and~\ref{fig:resonancespectra}(b), respectively) have XXX and XXX resonances between 0~G and 500~G, but only YYY and YYY for fields between 500~G and 1000~G. This is a direct consequence of the vibrational spacings being larger for more deeply bound vibrational states which give rise to Feshbach resonances at higher fields. 

In our analysis of the chaotic behaviour of resonance spacings we take into account the slow variation of resonance density by following the standard ``unfolding'' procedure~\cite{murpetit2015, green2016quantum}. Figures~\ref{fig:unfolding}(a)~and~(b) show a staircase function $S(B)=\sum_i \Theta(B_i)$, where $\Theta(x)$ is the Heaviside step function and $B_i$ are the resonance positions for the ``off-resonance'' and ``on-resonance'' datasets, respectively. $S(B)$ can be understood to give the total number of resonances for fields from zero up to $B$. We fit a smoothing spline, $\xi(B)$ to the staircase function. The function $\xi(B)$ is, by design, a continuous function of $B$ and can be viewed as a ``generalized'' resonance number with respect to $B$. Its values at resonance positions provide the unfolded resonance positions, $\xi_i = \xi(B_i)$. The derivative of $\xi$ with respect to the magnetic field, $\partial \xi/\partial B$ gives an estimate of the local resonance density, shown in Fig.~\ref{fig:unfolding}(e). In Er+Yb the resonance density for low fields $B<100~$G is slightly above $0.3$~G$^{-1}$, or approximately one resonance every 3~G. With higher fields the Feshbach spectrum becomes sparser, with $\partial \xi/\partial B \approx 0.1$~G$^{-1}$, or one resonance per 10~G, at about $B\approx 1000$~G. We find this behaviour to be the same for both the ``off-resonance'' and ``on-resonance'' cases.
% I fixed 
%whether the background scattering length $a_{\rm bg}$ is close to $\bar a$, as shown in Fig.~\ref{fig:unfolding}(b), or resonant.\todo{ PZ: we should not use the name "background scattering length" in this context: I would call it "zero field scattering length". Background scattering length is a property when the resonances are scarce and one can even define many background scattering lengths around indifidual resonances}

Figures~\ref{fig:unfolding}(c)~and~(d) show histograms of the nearest neighbor spacings $s_i = \xi_i-\xi_{i-1}$ for the Feshbach spectra in Figure~\ref{fig:resonancespectra}. If the resonances were randomly distributed, with no correlations, $s_i$ would exhibit a Poisson distribution, $P_{\rm P}(s) = \exp(-s)$. On the other hand, in the limit of strongly mixed states the resonances would tend to repel each other and the spacings would be described by a Wigner-Dyson distribution, \mbox{$P_{\rm WD}(s; \eta) = (\pi s/2)\exp(-\pi s^2/4)$},  where the probability density of two resonances overlapping is zero. Real systems can be described by the Brody distribution, $P_{\rm B}(s) = c_{\eta}(1+\eta)s^{\eta}\exp(-c_\eta s^{\eta+1})$, where $c_{\eta} = \Gamma\left((\eta+2)/(\eta+1)\right)^{\eta+1}$. The variable parameter $\eta$ makes it possible to smoothly interpolate between the idealized Poisson ($\eta=0$) and Wigner-Dyson ($\eta=1$) distributions. Following Green~\emph{et al.}~\cite{green2016quantum} we calculate the Brody parameter $\eta$ through max-likelihood estimation: $\eta$ can be found by maximizing the log-likelihood $l(\eta) = \sum_i \ln P_{\rm B}(s_i; \eta)$, while the standard deviation $u(\eta)=\left(\partial^2 l/\partial \eta^2 \right)^{-1/2}$. In practice, both $\eta$ and $u(\eta)$ can be easily found by fitting with a low order polynomial to a scan of $l(\eta)$ between $\eta=0$ and $\eta=1$. For the ``off-resonance'' case we find $\eta=0.71(11)$ and $\eta=0.54(10)$ for the ``on-resonance'' spectrum. These two values of $\eta$ are statistically consistent and characterize ErYb as a moderately chaotic system.

The number variance, $\Sigma^2(\Delta\xi)$, is defined as the variance of the number of resonances within the interval $[\xi_i, \xi_i+\Delta\xi]$, calculated over all resonances $\xi_i$. For a Poisson distribution of resonance spacings, $\Sigma^2_{\rm P}(\Delta\xi) = \Delta\xi$. For the Wigner-Dyson distribution the resonance spacings are much more even, and the number variance for large $\Delta\xi$ can be approximated with $\Sigma^2_{\rm WD}(\Delta\xi) = 2\pi^{-2}\left[\ln(2\pi\Delta\xi)+\gamma+1-\pi^2/8 \right]$, where $\gamma\approx0.577\ldots$ is the Euler constant. Figure~\ref{fig:unfolding}(f) shows the number variance calculated for the on- and  off-resonance cases concerned. We find that $\Sigma^2$ for the off-resonance case closely matches the Wigner-Dyson limit indicating strong level repulsion. The on-resonance result deviates slightly from the ideal Wigner-Dyson behaviour, but still indicates a strong signature of quantum chaos.

\section{Role of anisotropy\label{sec:anisotropy}}
%\todo{We should mention at some point that as the anisotropy is only from ab initio calculations we should bear in mind that in can be inaccurate - so we scan over anisotropy not only to better understand the mechanism of ememergence of chaos but also to see what happens if anisotropy is somewhat smaller/larger}
\begin{figure}%[!tbp]
    %\fbox{
    %\begin{minipage}{153mm}
    %    \flushright
    %    \vspace{6mm}
        \includegraphics[width=\textwidth]{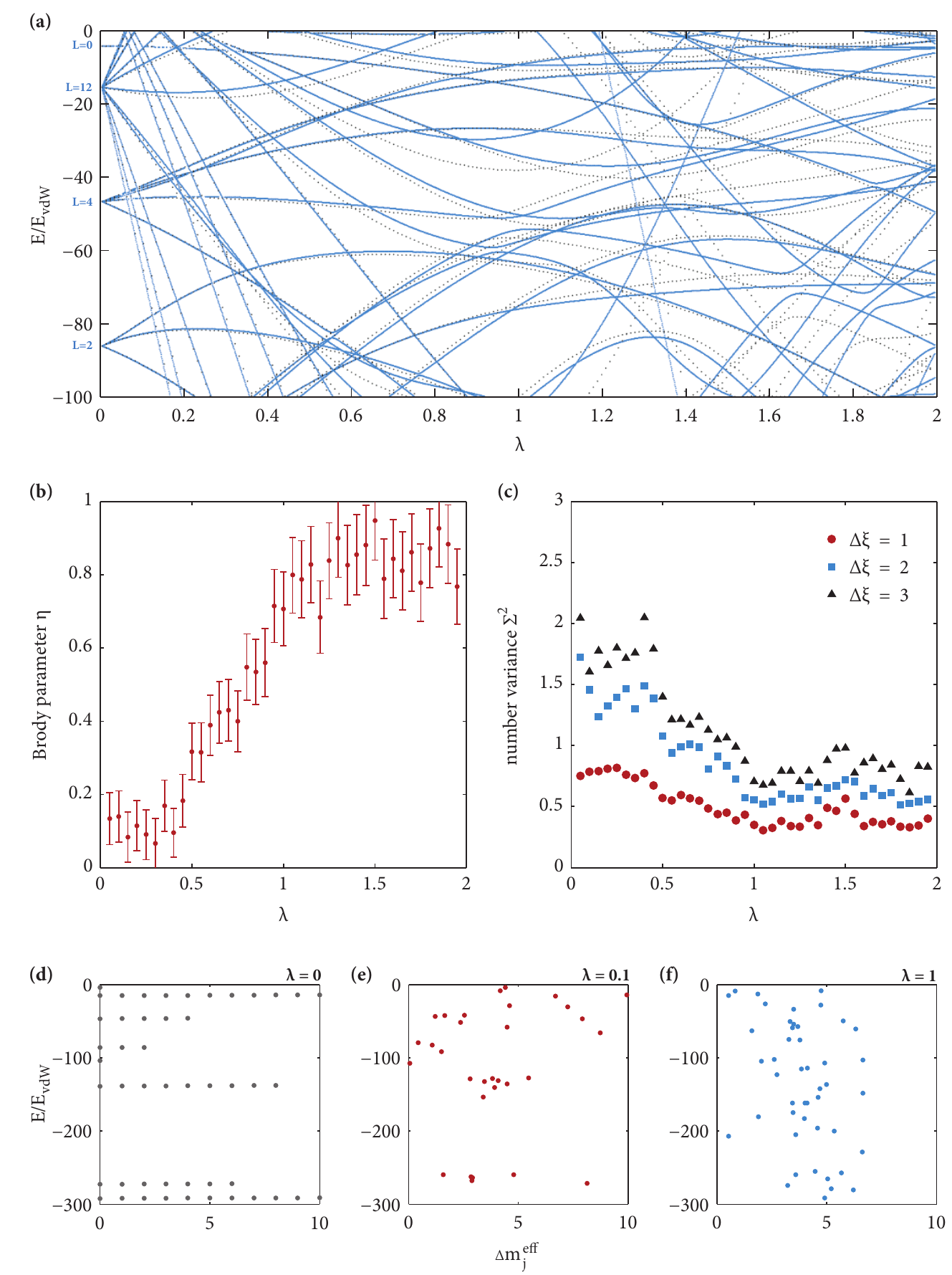}
        \caption{Emergence of chaotic physics as anisotropy is ramped up.  
                (a)~Near-threshold bound state energies as a function of an anisotropy scaling parameter, $\lambda$ (see text for details).
                (b)~Brody parameter $\eta$ as a function of $\lambda$. 
                (c)~Number variance $\Sigma^2$ shown for interval widths $\Delta\xi=1,\,2,\,3$. Here the emergence of Wigner-Dyson physics as anisotropy is ramped up is evidenced by the decrease in level number variance. 
                (d-f) Effective projection of the total electronic angular momentum, $m_j^{\rm eff}$, versus near-threshold bound state energies for $\lambda=0$~(d), $\lambda=0.1$~(e) and $\lambda=1$ (f). See Sections~\ref{sec:spectra}~and~\ref{sec:anisotropy} for details. 
   		}
        \label{fig:riseoftheanisotropy}
    %    \end{minipage}}
\end{figure}

The mechanism behind magnetic Feshbach resonances in ErYb is the anisotropy of atomic interactions that couples different even partial waves. To gain further insight into how chaos emerges in the Feshbach spectra we introduce an anisotropy scaling parameter $\lambda$, replace the anisotropic potentials ${\cal V}_{k=2,4,\ldots}$ by $\lambda {\cal V}_{k=2,4,\ldots}$ and repeat our statistical analysis as $\lambda$ is varied between zero and two. We should also point out that our knowledge of the anisotropy is of pure \emph{ab initio} origin and thus we also test the robustness of our predictions with respect to small inaccuracies in the total anisotropy. Figures~\ref{fig:riseoftheanisotropy}(a-f) show the previously discussed properties of the ErYb system as a function of $\lambda$ for the ``off-resonance case''.

We first look into zero-field near-threshold bound state energies. At zero anisotropy the vibrational levels have well defined rotational quantum numbers $L$ [solid lines in Figure~\ref{fig:riseoftheanisotropy}(a)]. As anisotropy is gradually introduced, the initial degeneracy of the $M_L$ sublevels is lifted and bound states split into $L+1$ components which correspond to states with different $|M_L|$ quantum numbers.
Interestingly, also quasibound states with $L>0$ become bound as the anisotropy emerges. 
%\sout{ with bound state energies varying in an approximately parabolic fashion }.% \todo{Nie ma zadnego parabolic fashion!}. 
At $\lambda \gtrapprox 0.8$ signatures of avoided crossings start to emerge and the system becomes manifestly chaotic for $\lambda \gtrapprox 1.1$. The unscaled ErYb system corresponds to $\lambda=1$ and this corroborates our previous conclusion that ErYb is a `moderately chaotic' system.

The statistical measures of chaos for our calculated Feshbach spectra also reveal the decisive impact of anisotropy on resonance correlations. Figure~\ref{fig:riseoftheanisotropy}(b) shows the Brody parameter $\eta$. At $\lambda$ close to zero, the Brody parameter $\eta \approx 0.1-0.3$ indicating a near complete lack of level repulsion. With $\lambda\approx1$ the values of $\eta$ are close to 0.7, again showing moderate chaos. For larger $\lambda$ the values of the Brody parameter appear to saturate near $\eta\approx 0.8-0.9$, consistent with strong level repulsion. Similarly, the number variances, shown in Figure~\ref{fig:riseoftheanisotropy}(c) for intervals $\Delta\xi=1,2,3$, all exhibit a downward trend as $\lambda$ is ramped.

We have also performed calculations of near-threshold bound states with the anisotropy limited to just the $V_2$ term [dotted lines in  Fig.~\ref{fig:riseoftheanisotropy}(a)]. Interestingly, the pattern of bound states in such case remains very close to one obtained with the full anisotropy (blue solid lines). Thus, it is clear that the role of interactions between closed channels is far more important than the direct coupling of $L=0$ scattering state with corresponding outgoing channels via $V_4$ and higher anisotropies. 
The dominating role of $V_2$ potential can be used in future for introducing a model potential that could reproduce the experimental spectrum of resonances using a small number of parameters. A similar situation can likely be expected in interactions of other systems with ultracold lanthanides: the most important information on the interaction could be hidden in the dominant anisotropic term, and the complex structure would originate from the coupling between closed channels.

The anisotropic potential $V_k$ mixes many channels with $M_L$ quantum number and $m_j$ which makes the bound states a composite of $m_j$ quantum states.
In the absence of anisotropy the differential magnetic moment with respect to the initial state $\Delta m_j^{\rm eff}$ (as defined in Sec. 4) takes even integer values between 2 and 12 [Fig. ~\ref{fig:riseoftheanisotropy}(d)]. As the anisotropy is ramped up $\Delta m_j^{\rm eff}$ becomes non-integer and starts from 0. Surprisingly, the mixing of $m_j$ even for a small anisotropy is quite strong: for $\lambda=0.1$ the $\Delta m_j^{\rm eff}$ bound states start to deviate from integer values [Fig. ~\ref{fig:riseoftheanisotropy}(e)]. For $\lambda=1$ the values of 
$\Delta m_j^{\rm eff}$ are completely mixed, with no discernible patterns [Fig. ~\ref{fig:riseoftheanisotropy}(f)].
 Since it is possible to experimentally measure the magnetic moments of the sub-threshold bound states by radio-frequency spectroscopy measurements~\cite{Frisch2015}, the statistical distribution of  $\Delta m_j^{\rm eff}$ could be investigated. The experimental results could be later confronted with theoretical predictions to test the anisotropy obtained with \emph{ab initio} calculations.  

\section{Sensitivity of Feshbach spectra to variations of reduced mass}

\begin{figure}%[!tbp]
    %\fbox{
    %\begin{minipage}{153mm}
    %    \flushright
    %    \vspace{6mm}
        \includegraphics[width=\textwidth]{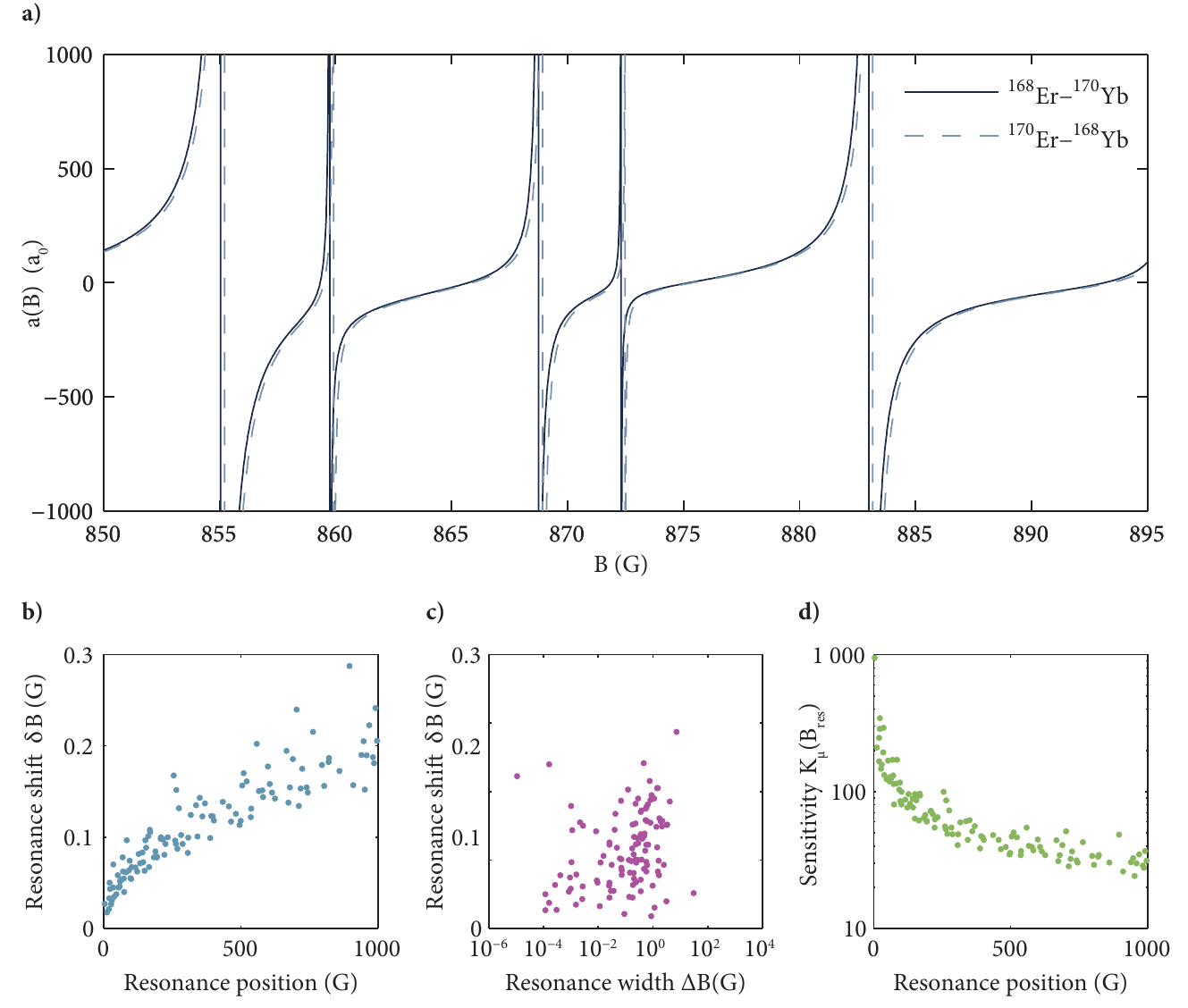}
        \caption{Sensitivity of magnetic Feshbach spectra to small changes in the reduced mass.  
                (a)~Comparison of Feshbach spectra for two isotopic combinations having nearly identical masses. Despite the chaotic spectrum, the mass change is small enough that resonance positions from one isotopic pair can be reliably matched to the other's.
                (b)~Resonance shifts vs resonance positions.
                (c)~Resonance shifts vs resonance widths.
                (d)~Relative sensitivity $K_\mu = (\mu/B)(\partial B/\partial \mu)$ to hypothetical changes to the proton-to-electron mass ratio as a function of resonance position.
   		}
        \label{fig:massratio}
    %    \end{minipage}}
\end{figure}

There is a number of pairs of ErYb isotopic combinations that have nearly identical reduced masses. In particular the smallest difference is between $^{168}$Er$^{170}$Yb and $^{170}$Er$^{168}$Yb, where the latter pair has a reduced mass larger by just $\delta \mu \approx 5.6\times 10^{-4}$~u (corresponding to a relative change $\delta \mu/\mu \approx 6.6\times 10^{-6}$). Given the minuscule change in the reduced mass, and the deterministic nature of the Schr\"odinger equations, one can expect the Feshbach spectra to be nearly identical for these two isotopic combinations. Indeed, the calculated spectra, shown in Fig.~\ref{fig:massratio}(a), are very similar, with resonance shifts much smaller than the spacings between resonances. This makes it possible to unambiguosly match the resonances between the two spectra despite the chaotic nature of the resonance spacing. The resonance shifts, shown in Fig.~\ref{fig:massratio}(b) vary roughly as $\delta B_{\rm res} \sim \sqrt{B_{\rm res}}$ and reach at most $0.3\,{\rm G}$ for resonance positions close to 1000~G. Resonances at larger fields have higher sensitivities to mass variations because they correspond to more deeply bound states, whose positions in turn are more sensitive~\cite{Chin2006, Zelevinsky2008}. The large variation in the shifts between adjacent resonances can be attributed to the repulsion between molecular bound states that is central to quantum chaos. We find no obvious correlations between the shifts and resonance widths $\Delta B$, as shown in Fig.~\ref{fig:massratio}(c).

Feshbach resonances have been proposed as a means for searching for temporal variations of the proton-to-electron mass ratio~\cite{Chin2006, Safronova2018}. In a molecular context, and in absence of hyperfine interactions, the sensitivity to this fundamental constant is mainly due to its direct influence on the reduced mass $\mu$. Our calculated shifts $\delta B_{\rm res}$ can thus be used to (trivially) evaluate the relative sensitivity 
\begin{equation}
    K_\mu (B_{\rm res}) = \frac{\mu}{B_{\rm res}} \frac{\partial B_{\rm res}}{\partial \mu}\,
\end{equation}
of resonance positions to variations in the reduced mass and, by proxy, the proton-to-electron mass ratio. The same could be done in experiment: a comparison of experimental Feshbach spectra for $^{168}$Er$^{170}$Yb and $^{170}$Er$^{168}$Yb could help choose the best (ie. most sensitive) Feshbach resonance to monitor for any variations. We note that both $^{168}$Yb and $^{170}$Yb naturally form stable Bose-Einstein condensates~\cite{Sugawa2011, Fukuhara2007} and any bosonic Er isotope can in principle be condensed because the Er-Er interaction can be tuned magnetically.

In ErYb $K_\mu (B_{\rm res})$ are largest for small fields, where for $B < 100\,{\rm G}$ we can expect many resonances with $K_\mu(B_{\rm res})$ of a few hundred. Given that magnetic fields can be controlled experimentally up to about $10^{-6}$, the positions of Feshbach resonances could constrain the variations of proton-to-electron mass ratio up to about $10^{-8}$---$10^{-9}$ at best. Much more useful constraints could potentially be achieved by monitoring the scattering length near a Feshbach resonance~\cite{Chin2006}. Assuming a Breit-Wigner profile [Eq.~(\ref{eq:breitwigner})] the relative sensitivity of the scattering length due to shift in resonance position\footnote{The contributions from $a_{\rm bg}$ and $\Delta B$ are negligible in comparison~\cite{Chin2006}.} is 
\begin{equation}
    K_\mu \left(a(B)\right) = \frac{\mu}{a(B)} \frac{\partial a(B)}{\partial \mu} 
    %= \frac{B_{\rm res}}{a(B)} \frac{\partial a(B)}{\partial B_{\rm res}} K_{\mu}(B_{\rm res}) 
    = \frac{B_{\rm res}\, \Delta B }{(B-B_{\rm res})(B-B_{\rm res}-\Delta B)} K_{\mu}(B_{\rm res}) \,.
\end{equation}
For realistic conditions of $B_{\rm res} = 100\,{\rm G}$, $\Delta B = 10^{-4}\,{\rm G}$, $K_{\mu}(B_{\rm res}) = 100$ and assuming a reasonable detuning from resonance, $B-B_{\rm res} = 2\Delta B$, we obtain $K_\mu \left(a(B)\right)$ on the order of $10^8$. If the scattering length can be measured to within $10^{-3}$---$10^{-6}$ then variations of the proton-to-electron mass ratio can be constrained down to $10^{-11}$---$10^{-14}$.

\section{Conclusion\label{sec:conclusion}}

In conclusion, we have investigated the ultracold collisions of bosonic Er and Yb atoms by means of accurate \emph{ab initio} calculations. Due to the slightly anisotropic Er atom, the ground state of ErYb is split into eleven asymptotically degenerate states with $\Lambda=0,\pm 1,\ldots,\pm 5$ with similar well depths, the deepest of which is 639~cm$^{-1}$. The potentials are split by at most 40~cm$^{-1}$ at equilibrium distance. The same small anisotropy ($C_{6,2}/C_{6,0} \approx 1\%$) is the mechanism behind the Feshbach resonances: the anisotropy couples different rotational quantum numbers. This coupling is also responsible for the emergence of chaos in the positions of both near-threshold bound states and magnetic Feshbach resonances. The resonance density is approximately $0.3\,{\rm G}^{-1}$ for small fields ($<100$~G) and decreases to about $0.1\,{\rm G}^{-1}$ at 1000~G. The resonance spacings exhibit features of quantum chaos as evidenced by a Brody parameter $\eta = 0.5-0.7$ and a number variance behaviour consistent with the predictions of the Wigner-Dyson theory. Despite the chaotic behavior we reliably observe several broad resonances for fields $<300\,{\rm G}$ that may be useful for scattering length control or molecular magnetoassociation. In our calculations using scaled anisotropy we have demonstrated the decisive impact of anisotropy on the emergence of quantum chaos in the Feshbach spectra. 

The ErYb system offers 20 bosonic isotopic combinations whose reduced masses range from 82.43~u (for $^{162}$Er$^{168}$Yb) to 85.92~u (for $^{170}$Er$^{176}$Yb) and cover a full cycle of scattering lengths. Uniquely, the pairs $^{168}$Er$^{170}$Yb and $^{170}$Er$^{168}$Yb have reduced masses that match to better than $10^{-5}$ relative. The resonance spectra for these pairs are so similar that it is possible to identify the resonances between them. In experiment this could be used to determine the sensitivity of these resonances to small variations of the proton-to-electron mass ratio with implications for fundamental physics~\cite{Chin2006, Zelevinsky2008, Safronova2018}. Experimentally, our predictions could be tested most cleanly by forming a dual Mott insulator of Yb and Er atoms. Separately, Mott insulators have already been demonstrated for both species. Keeping unity filling for the Er gas could help avoid additional losses due to the much denser Er-Er Feshbach spectrum. 

We expect our results for bosonic ErYb to qualitatively hold for other systems of highly magnetic lanthanides (like Dy, Ho, or Tm) combined with spin-singlet atoms (like Sr, Yb, or Hg). Since atoms in the $^1$S$_0$ are spherically symmetric, the anisotropy in a given combination of atomic species will only depend on the selected highly magnetic atom. Some variation between the different combinations may be expected due to the different reduced masses which in turn influence the rotational couplings. A combination of species involving the comparatively light Sr atom could form a mixture with a stronger mass imbalance that could be of interest in context of Efimov physics.

Lastly, ErYb offers a number of fermion-fermion and fermion-boson combinations which may be of interest for the quantum simulation and many-body physics communities. The behaviour of many-fermion ensembles is central to most important areas of physics, and in vastly different contexts: whether the design of new superconducting materials or understanding the structure of neutron stars. Finding ways to design new types of fermionic mixtures with various types of interactions (dipolar for Er and isotropic for Yb) can help build new types of experimental setups ready for new challenges in many-body quantum physics. Before the advent of quantum gases composed of lanthanides the research on fermionic mixtures was limited only to alkali quantum gases: potassium-40 and lithium-6, or their mixtures~\cite{Regal2004,Chin2004,Wille2008}. The Er+Yb system can make 2 Fermi-Fermi mixtures: $^{167}$Er combined with either the $^{171}$Yb or $^{173}$Yb isotope. The $^{167}$Er ground state has $f=j+s+i=19/2$. The first excited hyperfine state $f=17/2$ is separated by a 2.7~GHz gap~\cite{Frisch2013_hf}. Thus, the first crossing of the two lowest thresholds of $^{167}$Er+$^{171}$Yb and  $^{167}$Er+$^{173}$Yb systems alike corresponds to a magnetic field of $2.7$GHz$/g_j\mu_B=$1953~G. Close to the crossing the density of bound states supported by $f=17/2$ can be expected to be large leading to a densification of Feshbach resonance spectra for fields slightly below this value. An additional difference from the bosonic spectra is the possible Zeeman splitting of Feshbach resonances into multiplets due to the nuclear magnetic moment of fermionic Yb, similarly as in RbSr or CsYb molecules~\cite{Barbe2018,Yang2019}. The detailed analysis of interactions in fermion-fermion mixture, however, is beyond the scope of this paper, and we will study it in the near future.

\section*{Acknowledgments}

This research was supported by the National Science Centre (Grant No. 2017/25/B/ST4/01486).  
%Support has been received from project EMPIR 15SIB03 OC18. This project has received funding from the EMPIR programme co-financed by the Participating States and from the European Union’s Horizon 2020 research and innovation programme. 
This work is part of an ongoing research program at the National Laboratory FAMO in Toruń, Poland. 
Calculations have been carried out at the Wroclaw Centre for Networking and Supercomputing (http://www.wcss.pl), Grant Nos. 353 (MB) and 218 (PSŻ). PS\.Z is grateful to the FoKA society for encouragement. 

\section*{References}
\renewcommand{\section}[2]{}%

\bibliography{eryb}

\end{document}